\documentclass[twocolumn,showpacs,superscriptaddress,aps,prc,10pt,nofootinbib]{revtex4-1} 
\usepackage[pdftex]{graphicx} 
\usepackage{cancel} 
\usepackage[usenames,svgnames,table]{xcolor} 
\usepackage[english]{babel} 
\usepackage{ucs} 
\usepackage[utf8]{inputenc} 
\usepackage{siunitx} 
\usepackage[version=3]{mhchem} 
\usepackage{multirow} 
\usepackage{textcomp} 
\usepackage{soul} 
\usepackage{dcolumn}
\usepackage{bm}
\newcommand{\faz}{\textsc{FAZIA}}

\newcommand{\gem}{\textsc{Gemini++}} 
\newcommand{\ca}{$^{40}$Ca} 
 
\newcommand{\caa}{$^{48}$Ca} 

\newcommand {\amev}{$\,$MeV/nucleon}

\newcommand {\qpr}{QP$_{R}$}
\newcommand {\qpe}{QP$_{E}$}
\newcommand {\qpb}{QP$_{B}$}
\newcommand {\dee}{$\Delta$E-E}
\newcommand {\vs}{\textit{vs.}}
\newcommand {\ie}{\textit{i.e.}}

\newcommand {\pred}{$p_{red}$}
\newcommand {\bred}{$b_{red}$}
\newcommand {\bredb}{$\langle b_{red} \rangle$}
\newcommand {\neos}{nEoS}

\newcommand {\nsuz}{$\langle N/Z \rangle$}

\usepackage[hyperfootnotes=false,colorlinks=true,allcolors=blue!60!black]{hyperref}

\begin{document} 
%
%
\title{Isospin diffusion measurement \\from the direct detection of a Quasi-Projectile remnant}

\author{A. Camaiani} \email{alberto.camaiani@fi.infn.it} 
\affiliation{Dipartimento di Fisica, Universit\`a di Firenze, Italy} 
\affiliation{INFN, Sezione di Firenze, Italy} 
\author{G. Casini} 
\affiliation{INFN, Sezione di Firenze, Italy} 
\author{S. Piantelli} 
\affiliation{INFN, Sezione di Firenze, Italy} 
\author{A. Ono}
\affiliation{Department of Physics, Tohoku University, Sendai 980-8578, Japan}
\author{E. Bonnet}
\affiliation{SUBATECH, Université de Nantes, IMT Atlantique, IN2P3/CNRS, 4 Rue Alfred Kastler, 44307 Nantes Cedex 3, France} 
\author{R. Alba}
\affiliation{INFN Laboratori Nazionali del Sud, Via S. Sofia 62, 95125 Catania, Italy}
\author{S. Barlini}
\affiliation{Dipartimento di Fisica, Universit\`a di Firenze, Italy} 
\affiliation{INFN, Sezione di Firenze, Italy} 
\author {B. Borderie}
  \affiliation{Universit\'e Paris-Saclay, CNRS/IN2P3, IJCLab, 91405 Orsay, France}
  \author{R. Bougault}
  \affiliation{Normandie Universit\'e, ENSICAEN, UNICAEN, CNRS/IN2P3, LPC Caen, 14000 Caen, France}
  \author{C. Ciampi}
  \affiliation{Dipartimento di Fisica,  Universit\`a di Firenze, Italy} 
  \author{A. Chbihi}
  \affiliation{Grand Accélérateur National d’Ions Lourds (GANIL), CEA/DRF - CNRS/IN2P3, Boulevard Henri Becquerel, F-14076 Caen, France}
  \author{M. Cicerchia}
  \affiliation{INFN Laboratori Nazionali di Legnaro, 35020 Legnaro, Italy}
  \author{M. Cinausero}
  \affiliation{INFN Laboratori Nazionali di Legnaro, 35020 Legnaro, Italy}
\author{J.A. Due\~nas}
\affiliation{Depto. de Ingenier\'ia El\'ectrica y Centro de Estudios Avanzados en F\'isica, Matem\'aticas y
Computaci\'on, Universidad de Huelva, 21007 Huelva, Spain } 
\author{D. Dell'Aquila}
\affiliation{Dipartimento di Chimica e Farmacia, Università degli Studi di Sassari, Sassari, Italy}
\affiliation{INFN - Laboratori Nazionali del Sud, Catania, Italy}
\author{Q. Fable}
\affiliation{Normandie Universit\'e, ENSICAEN, UNICAEN, CNRS/IN2P3, LPC Caen, 14000 Caen, France}
\author{D. Fabris}
\affiliation{INFN Sezione di Padova, 35131 Padova, Italy}
\author{C. Frosin} 
  \affiliation{Dipartimento di Fisica,  Universit\`a di Firenze, Italy} 
  \affiliation{INFN, Sezione di Firenze, Italy} 
  \author{J. D. Frankland}
  \affiliation{Grand Accélérateur National d’Ions Lourds (GANIL), CEA/DRF - CNRS/IN2P3, Boulevard Henri Becquerel, F-14076 Caen, France}
\author{F. Gramegna}
\affiliation{INFN Laboratori Nazionali di Legnaro, 35020 Legnaro, Italy}
 \author{D. Gruyer}
  \affiliation{Normandie Universit\'e, ENSICAEN, UNICAEN, CNRS/IN2P3, LPC Caen, 14000 Caen, France}
\author{K. I. Hahn}
\affiliation{Department of Science Education, Ewha Womans University, Seoul 03760, Republic of Korea}  
\author{M. Henri}
\affiliation{Grand Accélérateur National d’Ions Lourds (GANIL), CEA/DRF - CNRS/IN2P3, Boulevard Henri Becquerel, F-14076 Caen, France}
\author{B. Hong}
\affiliation{Center for Extreme Nuclear Matters (CENuM), Korea University, Seoul 02841, Republic of Korea}
\affiliation{Department of Physics,  Korea University, Seoul 02841, Republic of Korea}
\author{S. Kim}
\affiliation{Department of Science Education, Ewha Womans University, Seoul 03760, Republic of Korea }
\author{A. Kordyasz}
\affiliation{Heavy Ion Laboratory, University of Warsaw, 02-093 Warszawa, Poland}
\author{M. J. Kweon}
\affiliation{Center for Extreme Nuclear Matters (CENuM), Korea University, Seoul 02841, Republic of Korea}
\affiliation{Department of Physics,  Inha University, Incheon 22212, Republic of Korea}
\author{H. J. Lee}
\affiliation{Department of Physics,  Inha University, Incheon 22212, Republic of Korea}
\author{J. Lemari\'e}
\affiliation{Grand Accélérateur National d’Ions Lourds (GANIL), CEA/DRF - CNRS/IN2P3, Boulevard Henri Becquerel, F-14076 Caen, France}
  \author{N. LeNeindre}
  \affiliation{Normandie Universit\'e, ENSICAEN, UNICAEN, CNRS/IN2P3, LPC Caen, 14000 Caen, France}
  \author{I. Lombardo}
  \affiliation{INFN Sezione di Catania, 95123 Catania, Italy} 
\author{O. Lopez}
  \affiliation{Normandie Universit\'e, ENSICAEN, UNICAEN, CNRS/IN2P3, LPC Caen, 14000 Caen, France}
\author{T. Marchi}
\affiliation{INFN Laboratori Nazionali di Legnaro, 35020 Legnaro, Italy}

\author{S. H. Nam}
\affiliation{Center for Extreme Nuclear Matters (CENuM), Korea University, Seoul 02841, Republic of Korea}
\affiliation{Department of Physics,  Korea University, Seoul 02841, Republic of Korea}

 \author{P. Ottanelli}
  \affiliation{Dipartimento di Fisica,  Universit\`a di Firenze, Italy} 
  \affiliation{INFN, Sezione di Firenze, Italy}   
\author{M. Parlog}
  \affiliation{Normandie Universit\'e, ENSICAEN, UNICAEN, CNRS/IN2P3, LPC Caen, 14000 Caen, France}
  \affiliation{"Horia Hulubei" National Institute of Physics and Nuclear Engineering (IFIN-HH), RO-077125 Bucharest Magurele, Romania}
  \author{G. Pasquali}
\affiliation{Dipartimento di Fisica,  Universit\`a di Firenze, Italy} 
  \affiliation{INFN, Sezione di Firenze, Italy} 
\author{G. Poggi}
\affiliation{Dipartimento di Fisica, Universit\`a di Firenze, Italy} 
\affiliation{INFN, Sezione di Firenze, Italy} 
\author{J. Quicray}
\affiliation{Normandie Universit\'e, ENSICAEN, UNICAEN, CNRS/IN2P3, LPC Caen, 14000 Caen, France}
\author{A. A. Stefanini}
\affiliation{Dipartimento di Fisica, Universit\`a di Firenze, Italy} 
\affiliation{INFN, Sezione di Firenze, Italy} 
\author{S. Upadhyaya}
\affiliation{Faculty of Physics, Astronomy and Applied Computer Science, Jagiellonian University, 30-348 Kracow, Poland}
\author{S. Valdr\'e}
\affiliation{INFN, Sezione di Firenze, Italy}
\author{E. Vient}
\affiliation{Normandie Universit\'e, ENSICAEN, UNICAEN, CNRS/IN2P3, LPC Caen, 14000 Caen, France}

\begin{abstract} 

The neutron-proton equilibration process in \caa{}+\ca{} at 35\amev{} bombarding energy has been experimentally estimated by means 
of the isospin transport ratio. Experimental data have been collected with a subset of the FAZIA telescope array, which permitted to determine $Z$ and $N$ of detected fragments. For the first time, the QP evaporative channel has been compared with the QP break-up one in a homogeneous and consistent way, pointing out to a comparable n-p equilibration which suggests close interaction time between projectile and target independently of the exit channel. Moreover, in the QP evaporative channel n-p equilibration has been compared with the prediction of the Antisymmetrized Molecular Dynamics (AMD) model coupled to the GEMINI statistical model as an afterburner, showing a larger probability of proton and neutron transfers in the simulation with respect to the experimental data.  
\end{abstract}

\maketitle 

\section{Introduction}

Since the end of the '80s some experiments, mostly focused 
on dissipative collisions below 20\amev{}, investigated how a colliding system with projectile 
and target with different ``chemical'' composition, evolves towards the charge equilibration~\cite{bib:planeta, bib:gippner, bib:madani, bib:marchetti}. Later on, the so-called isospin dynamics, namely 
the neutron-proton (n-p) exchange between two interacting nuclei,
gained much attention at Fermi energies (20-100\amev{}), 
where nuclear subsystems relatively far from the saturation value of the baryon density can be explored; this, in turn, allows to investigate how the nuclear Equation of State (\neos{}) rules the dynamics~\cite{bib:betty04_isoscaling, bib:betty09_isoscaling}.
In the Fermi energy domain, interesting signals have been found mainly in binary semi-peripheral collisions, mosrly the clear evidence of a neutron enrichment of the fragments emitted from the phase-space region between the two main reaction products (also labeled mid-velocity or neck region)~\cite{bib:lukasik95, bib:plagnol00_lcp, bib:theriault05, bib:theriault06}.
A theoretical interpretation was proposed and timely developed in the framework of nuclear reaction models, in order to describe the isotopic composition of the emerging excited Quasi-Projectile (QP) and Quasi-Target (QT) after the collision: the n-p equilibration is largely due to the initial different concentration of neutrons and protons between projectile and target (isospin diffusion) while the neutron enrichment of the mid-velocity zone is ascribed to the density gradient, which arises between the different regions of the colliding systems (isospin drift)~\cite{bib:baran05_eos, bib:lionti05_imf, bib:napo10_eos}. In this paper, we discuss about isospin diffusion and how it guides the system towards the n-p equilibration. 

The degree of charge equilibration is strictly related both to the driving force which rules the n-p exchange and to the interaction time. In particular, the isospin diffusion is sensitive to the symmetry energy term $E_{\mathrm{sym}}$ of the \neos{}~\cite{bib:napo10_eos, bib:baran05_eos}, and it has been used, in the past, to put some constraints on that and on the whole parametrization~\cite{bib:betty04_isoscaling, bib:betty09_isoscaling, bib:sun10_isobaricratio}. However, to date, a clear knowledge of the symmetry energy is still lacking, namely the Taylor expansion coefficients are known with large uncertainties (first order term, $L_{\mathrm{sym}}$) or not at all (second order, $K_{\mathrm{sym}}$, and higher order coefficients)~\cite{bib:magueron18_eos}. Concerning the interaction time, for a given restoring potential, the longer the interaction time the more equilibrated in isospin the system~\cite{bib:betty04_isoscaling}. In this sense, different effects contribute to the equilibration, such as in-medium effects which significantly reduce the nucleon-nucleon cross section with respect to the nucleon-nucleon value~\cite{bib:lopez14_inmedium}, or cluster correlations that arise during the collision~\cite{bib:coupland11_imbalance_effects}. Therefore, a characterization of the collision as a function of the reaction centrality is mandatory in order to explore different interaction times. 

During the years the experimental investigations followed two 
main paths. The first one exploited detection arrays covering a large part of the solid angle in order to globally characterize the acquired events, although with limitations in terms of isotopic separation (typically below $Z\approx 8$)~\cite{bib:miniball, bib:indra, bib:pagano12_chimera2}. As a consequence, in such studies~\cite{bib:betty04_isoscaling, bib:betty09_isoscaling, bib:txliu07_isobaricratio, bib:galichet09_qprebuild, bib:sun10_isobaricratio, bib:bougault18_emission} only the lightest QP decay products could be used to extract information on the isospin equilibration. The second one adopted mass spectrometers, in order to directly access to the neutron-proton ratio ($N/Z$) of the QP remnants, at the expense of covering a small part of the solid angle and detecting only the main fragment of the event. Consequently, no information on break-up events or Intermediate Mass Fragments (IMFs) and/or Light Charged Particles (LCPs) accompanying the QP could be obtained in typical configurations~\cite{bib:souliotis06_qpr, bib:souliotis14_qpr}. On the other hand, according to the literature~\cite{bib:baran05_transport, bib:napo10_eos}, the experimental determination of the $N/Z$ content of the QP remnant could be a good probe to put constraints on the symmetry energy. 
In such a scenario, it could be useful to directly detect the isospin content of the QP remnant, together with the accompanying particles or fragments. An example in this direction is the recent paper of the NIMROD collaboration where the authors reconstructed the isospin of the QP remnant~\cite{bib:may19_ratio}.

The present work fits with this panorama, aiming at the investigation 
of the isospin diffusion in peripheral and semi-peripheral reactions and trying to overcome the limitation of previous detectors. 
In fact we investigated the asymmetric reaction \caa{}+\ca{} at 35\amev{} by means of the \faz{} multi-telescope array, mainly for two reasons. Firstly, Ca isotopes allow to stress the isospin unbalance of the entrance channel, moving from $(\frac{N}{Z})_{^{48}Ca}=1.4$ to $(\frac{N}{Z})_{^{40}Ca}=1$. Secondly, for such reactions the \faz{} array allows a mass resolution comparable with that of a spectrometer~\cite{bib:bougault14}, allowing to fully access the isotopic content of the QP remnant. Moreover, thanks to the good granularity of the detector, we can investigate also the break-up channel in order to isotopically reconstruct the QP from the detected pair~\cite{bib:camaiani18_iwm}. In light of this, we measured the n-p equilibration in the QP evaporative channel, directly accessing the QP remnant; this will be compared for the first time, in a homogeneous and coherent way, with the QP break-up channel, where the QP can be reconstructed from the daughter fragments.

In order to extract the equilibration degree in \caa{}+\ca{} system, referred in the following as the mixed one, we adopted the isospin transport ratio (also known as imbalance ratio)~\cite{bib:rami00_imbalance}, which normalizes 
an isospin related observable measured in the asymmetric system to that measured for two symmetric reactions, where the isospin diffusion is absent by definition. For this reason, \caa{}+\caa{} and \ca{}+\ca{} reactions, both at 35\amev{}, have been also measured and used as reference. The isospin transport ratio is defined as follows~\cite{bib:rami00_imbalance}: 
\begin{equation}
R(X) = \frac{2X - X^{4848} - X^{4040}}{X^{4848} - X^{4040}}
\label{eq:ratio}
\end{equation}
where $X$ is an isospin sensitive observable evaluated for the three systems. 
For the two symmetric systems \caa{}+\caa{} and \ca{}+\ca{}, $R(X)$  assumes the value of +1 and -1, respectively. Such a method allows to enhance the equilibration signal due to the isospin diffusion~\cite{bib:betty04_isoscaling, bib:txliu07_isobaricratio, bib:may19_ratio}, reducing the effects of any unwanted overlapping process, and effectively cancelling those introducing a linear transformation of $X$~\cite{bib:camaiani20_ratio}. Moreover, we note that if the chosen variable linearly depends on the isospin of the system, $R(X) = \pm 1$ represents the ``No Equilibration'' limit, while $R(X) = 0$ the ``Full Equilibration'' value~\cite{bib:rami00_imbalance}. As done in the past~\cite{bib:planeta, bib:gippner, bib:madani, bib:marchetti}, in this paper, the n-p equilibration is followed as a function of the reaction dissipation. Since the impact parameter is not directly accessible as an experimental observable, as usual, we used a reaction centrality estimator whose effectiveness to follow the impact parameter order has been
tested by means of the Antisymmetrized Molecular Dynamics (AMD)~\cite{bib:amd92} model coupled with GEMINI++~\cite{bib:gemini} as an afterburner.

This paper is organized as follows. 
In Section~\ref{sec:exp} the experimental apparatus
and the adopted theoretical models are presented. 
Section~\ref{sec:evselandgross} describes the event selection criteria; also the gross properties of the studied systems are presented. The adopted method to estimate the reaction centrality is presented in Section~\ref{sec:dissipation}.
The n-p equilibration in both the QP evaporative and QP break-up channels is presented in Section~\ref{sec:eq}, while the comparison of the QP evaporative channel with the AMD+GEMINI++ prediction is reported in Section~\ref{sec:amd}. Summary and conclusions are given in Section~\ref{sec:concl}.

\section{Investigation approach}
\label{sec:exp}

We performed the experiment using beams of $^{40,48}$Ca at 35\amev{}, delivered by the Superconducting Cyclotron of INFN-LNS with an average current of 0.1$\,$pnA, impinging on $^{40,48}$Ca targets with a thickness of 500$\,\mu$g/cm$^{2}$. Approximately 110, 70 and 15 millions of events have been collected for the \caa{}+\caa{}, \caa{}+\ca{} and \ca{}+\ca{}, respectively. The vacuum inside the scattering chamber was 2$\times$10$^{-5}\,$mbar during the whole experiment. 

In order to avoid Ca oxidation during the mounting of the targets, the Ca layers were sandwiched between two Carbon foils of about 10$\,\mu$g/cm$^{2}$ on both sides of each target was used. Data of both $^{40,48}$Ca beams impinging on $^{12}$C (300$\,\mu$g/cm$^{2}$ thick) have been collected in order to estimate the carbon reaction background in the main reaction data. As observed in a previous analysis where the same Ca targets have been used~\cite{bib:piantelli19_isofazia}, no significant contribution of reactions on Carbon target has been found, thus concluding that the background due to reaction on Carbon negligibly affects the present results~\cite{bib:phd_camaiani}.

\begin{figure}
\centering 
\includegraphics[width=0.8\columnwidth]{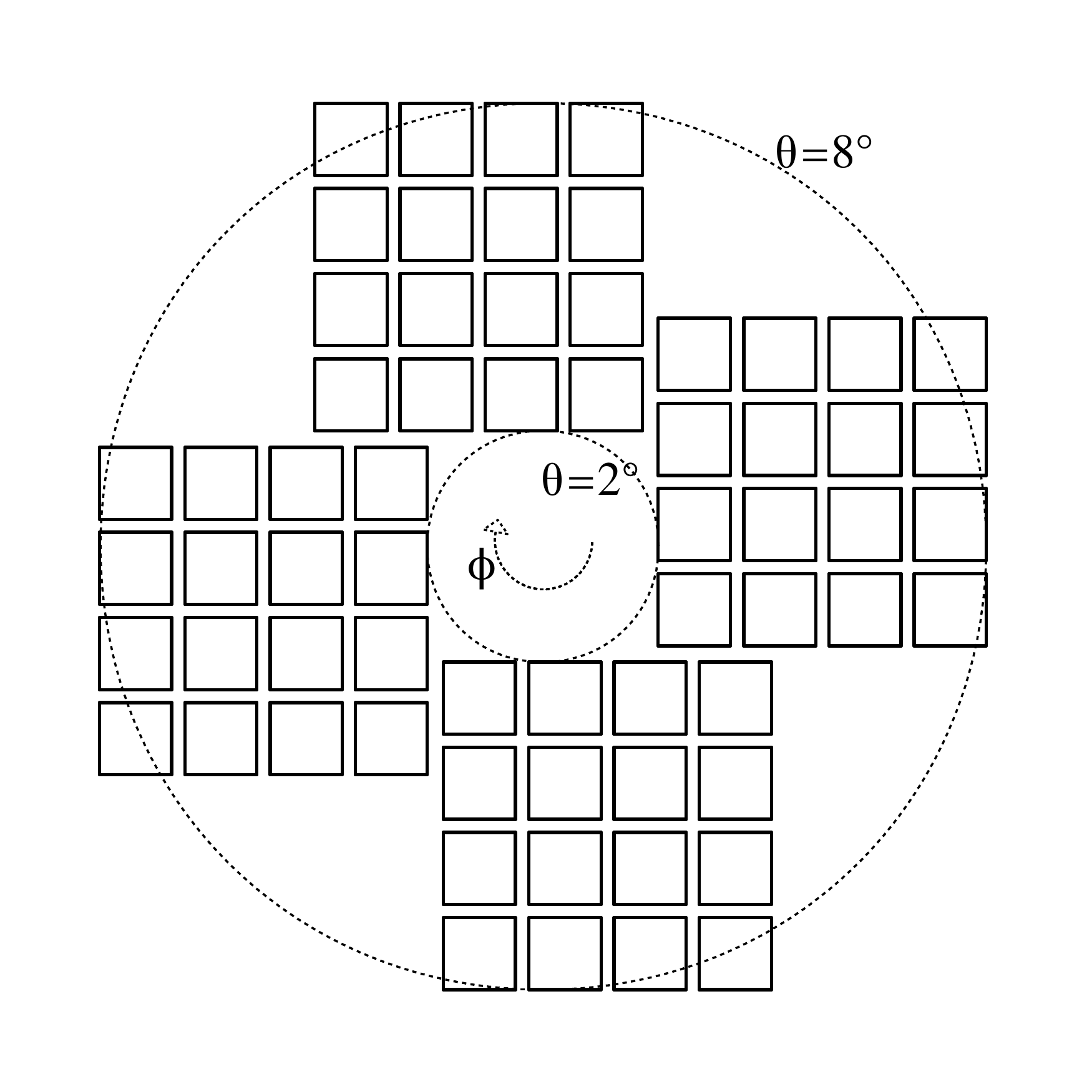} 
\caption{Schematic polar representation of the apparatus geometry. The beam axis passes through the symmetry center. View from the target.} 
\label{fig:schema} 
\end{figure}

Data have been collected with four \faz{} blocks~\cite{bib:bougault14, bib:valdre18} arranged in a wall configuration around the beam axis covering polar angles from 2$^{\circ}$ up to 8$^{\circ}$ approximately, 80$\,$cm far from the target. A schematic representation of the apparatus geometry is shown in Fig.~\ref{fig:schema}.
The main features and performances of the \faz{} multi-telescope array are fully described elsewhere~\cite{bib:bougault14, bib:valdre18, bib:pastore_psa, bib:frosin19_csi}.
Here, we remind that each block consists of 16 2$\times$2$\,$cm$^{2}$ Si-Si-CsI(Tl) telescopes, 
where the thickness of different layers is 300$\,\mu$m, 500$\,\mu$m, and 10$\,$cm, respectively. The telescopes are directly coupled to ``custom'' FEE cards, featuring the preamplifiers and the fast digital sampling stages, also allowing the on-line extraction of the energy parameters from the signals~\cite{bib:valdre18}. Each \faz{} telescope allows to identify iostopes in charge and mass up to $Z\approx25$ with the \dee{} technique~\cite{bib:carboni_psa} and up to $Z\approx20$ via Pulse Shape Analysis in Silicon detectors~\cite{bib:pastore_psa} for fragments stopped in the first Silicon layer with identification energy threshold depending on the ion charge~\cite{bib:pastore_psa}. The data presented in this paper refer to the QP phase-space; as in most other experiments, energy thresholds do not allow to access the QT phase-space, which results almost undetected.

\begin{figure*}
\centering 
\includegraphics[width=1\textwidth]{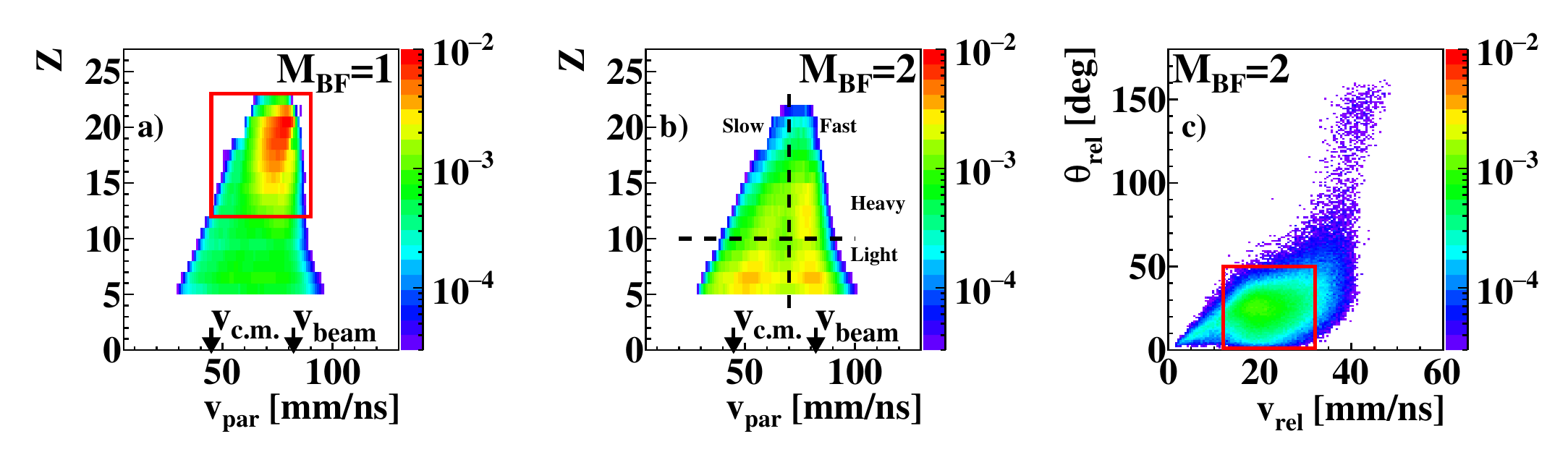} 
\caption{(Color online) Experimental data for the \caa{}+\caa{} reaction. Panel a-b) Charge \vs{} parallel velocity correlation in laboratory frame of $BF$ ejectiles. a): events with $M_{BF}=1$, the rectangle shows the \qpr{} selection; b) events with $M_{BF}=2$. Beam ($v_{beam}$) and c.m. system ($v_{cm}$) velocity are pointed out by the arrows. Panel c) $\theta_{rel}$ \vs{} $v_{rel}$ correlation between the two $BF$s of the same events as in panel b); the rectangle points out the \qpb{} selection. 
Each correlation is normalized to unitary integral.} 
\label{fig:hsel_zv} 
\end{figure*}

As anticipated, from the theoretical side, data are compared with the predictions of the AMD model, belonging to the Quantum Molecular Dynamics family~\cite{bib:aichelin86_qmd1, bib:aichelin91_qmd2}, due to its well assessed capability to describe nuclear collision characteristics in a various range of energy and impact parameters~\cite{bib:ono19_rev}. 
In brief, this model describes a many-body nuclear system by means 
of a Slater determinant of Gaussian wave packets and the equation of motion is obtained via time dependent variational principle~\cite{bib:amd16_pion}. 
The version of the AMD code used in this work implements the mean-field via the effective interaction Skyrme SLy4~\cite{bib:skyrme}, using $K_{sat}=230\,$MeV for the incompressibility modulus of the nuclear-matter and $\rho_0 = 0.16\,$fm$^{-3}$ for the saturation density. Two parametrizations of the symmetry energy can be tested within AMD: an asym-soft one with $E_{sym}=32\,$MeV and $L_{sym}=46\,$MeV, and an asym-stiff one with $L_{sym}=108\,$MeV and the same value for $E_{sym}$, obtained by changing the density dependent term in the SLy4 force~\cite{bib:amd16_pion}. Such recipes are compatible with
the reported values for realistic parametrizations~\cite{bib:magueron18_eos}.
Nucleon-nucleon collisions are taken into account by implementing test particles which are randomly generated at every time step~\cite{bib:ono19_rev, bib:piantelli19_fiasco}.
The transition probability depends on the in-medium nucleon-nucleon cross section, which can be considered, within some limits, as a free parameter of the model. In this used code version, the parametrization proposed in Ref.~\cite{bib:coupland11_imbalance_effects} has been used, \ie{} 
$\sigma = \sigma_{0} \tanh \left( \sigma_{\mathrm{free}} / \sigma_{0} \right)$,
with $\sigma_{0}= y\rho^{-2/3}$ , where $y$ is a screening parameter, set at $y = 0.85$ (according to ~\cite{bib:coupland11_imbalance_effects}).  In order to
take into account cluster correlations arising during the
dynamics, cluster states are included among the possible
achievable final states~\cite{bib:tian17_cc, bib:tian18_cc, bib:ono19_rev, bib:piantelli19_fiasco}.

We produced about 40000 events for each system and symmetry energy parametrization, 
stopping the dynamical calculation at 500$\,$fm/c, a time when the dynamical phase is safely concluded and the Coulomb interaction among QP and QT can be considered negligible~\cite{bib:piantelli19_fiasco}. Impact parameters up to the grazing values $b_{gr}$ (10.4, 10.1 and 9.7$\,$fm for the n-rich, mixed and n-deficient system, respectively) have been randomly sorted, with a triangular distribution. For each primary event, 2000 secondary events have been generated by means of the GEMINI++~\cite{bib:gemini} statistical Monte Carlo code. The simulated data have then been filtered through a software replica of the apparatus, that takes into account the geometrical efficiency and the identification thresholds, in order to consistently compare the simulation output with the experimental results.

\section{Event selection and reaction characterization}
\label{sec:evselandgross}

In order to show the criteria adopted for selecting events we focus on the \caa{}+\caa{} reaction for the sake of brevity. 
The same selection criteria have been applied to the other systems. First of all, due to pile-up events,  events with the total detected charge $Z_{TOT}$ greater than the total system charge are rejected, as well as events with a total parallel momentum greater than the beam momentum (less than 2\%). Only events with isotopically identified ejectiles have been considered in the present work, which represents more than 80\% of the total events. 

The event selection is based on a detected multiplicity ($M$) condition. We define as Big Fragments ($BF$s), any ejectile with $Z\geq5$, and as IMFs only Lithium and Beryllium ions.
This choice is motivated by the fact that most particles with $Z<5$ come from statistical emission according to the AMD+GEMINI++ predictions.
According to our goal, we want to select two main channels, \ie{} the evaporative channel, and the break-up one. In the evaporative channel the primary QP de-excites emitting IMFs and LCPs, thus only a BF is expected. Differently, in the break-up channel, the primary QP splits in two BFs, possibly excited above the energy threshold for particle decay and thus undergoing subsequent evaporation. Consequently, the first class is identified by the presence of one $BF$ ($M_{BF}=1$), while the second one includes two $BF$ ($M_{BF}=2$). It is worth mentioning that these classes correspond to 65\% and 2\% of the total number of acquired events, respectively; the remaining part, due to the limited solid angle coverage, contains events with only LCP and/or IMF detected and it is discarded.

\begin{figure*}
\centering 
\includegraphics[width=1\textwidth]{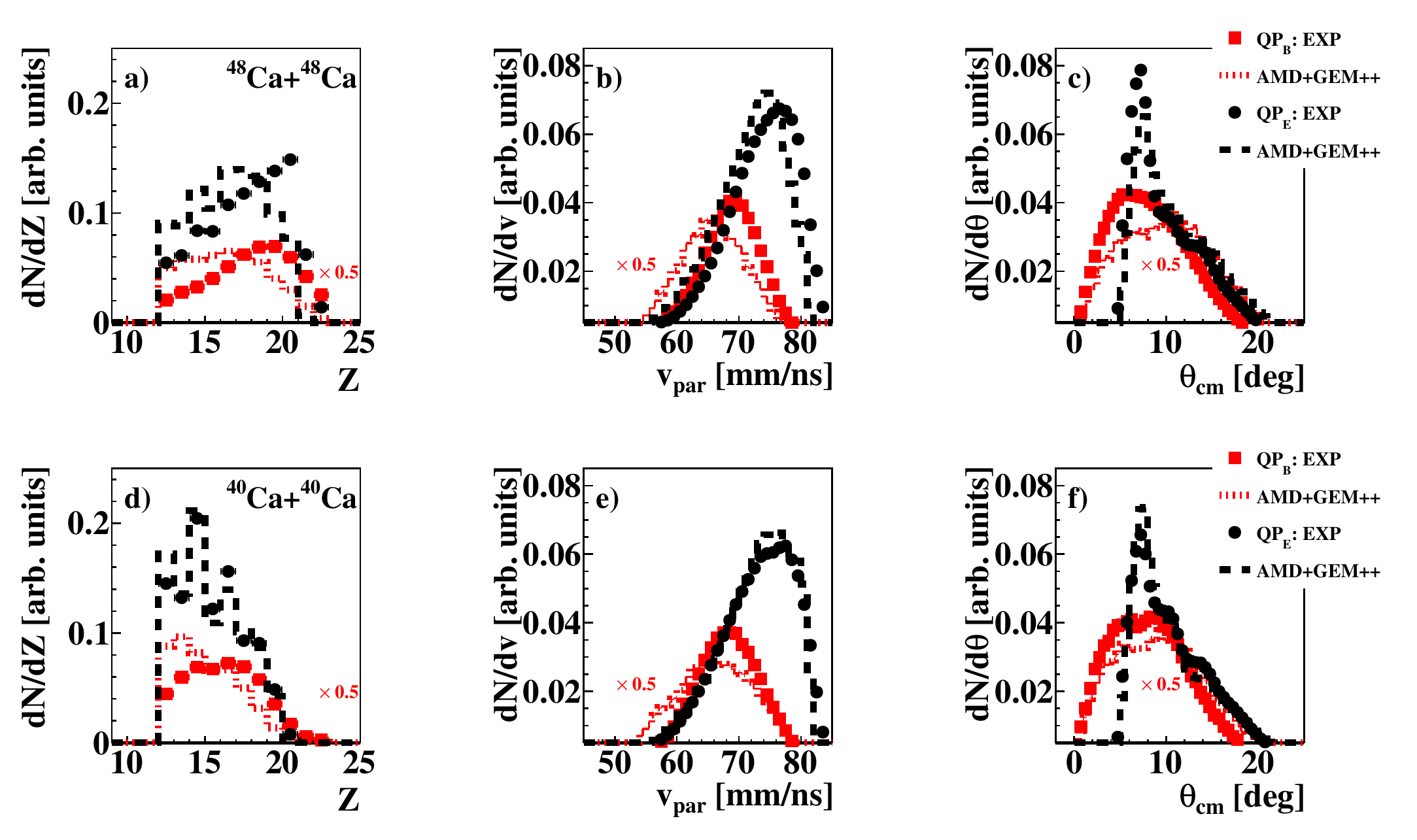} 
\caption{(Color online) 
Experimental (symbols) and simulated (lines) properties of the QP in the \qpe{} (black) and \qpb{} (red) channels, for the \caa{}+\caa{} (a-c) and \ca{}+\ca{} (d-f) reactions. Panel a,d): charge distributions. Panel b,e): parallel velocity in the laboratory frame. Panel c,f): polar angle in the c.m. system. Each distribution is normalized to unitary integral. \qpb{} distributions are scaled by a factor 0.5 for sake of clarity. Statistical errors are smaller than the marker size (line width).} 
\label{fig:hsel_gross} 
\end{figure*} 

Fig.~\ref{fig:hsel_zv}(a,b) shows the $BF$ charge \vs{} the parallel velocity (along the beam axis, $v_{par}$) correlation in the laboratory frame for events with $M_{BF}=1$ and $M_{BF}=2$, respectively. 
Beam ($v_{beam}$) and center of mass ($v_{cm}$) velocities are pointed out by the arrows. Panel a) shows a quite intense spot in the charge region 
$12 \leq Z \leq 22$, with parallel velocity between 60 and 80$\,$mm/ns (\ie{} $BF$s that preserve down to the 75\% of the projectile velocity). The $BF$s whose charge is greater than the projectile charge are ascribable to a charge transfer from the target to the projectile during the interaction phase. Both charge and velocity are compatible with a $BF$ that is the QP remnant after the de-excitation through the emission of LCP and/or IMF.  The observed spot corresponds to a projectile that retains down to 60\% of its initial charge: such charge range complies with analogous selections adopted in literature~\cite{bib:galichet09_qprebuild, bib:galanopoulos10_fission}. 
As a consequence, we select as QP evaporative channel (\qpe{}) those events containing a QP remnant (labeled as \qpr{}), \ie{} a $BF$ forward emitted with $Z = 12 \div 22$), as pointed out by the red contour in Fig.~\ref{fig:hsel_zv}(a). \qpe{} events represent 52\% of the total collected data.

Fig.~\ref{fig:hsel_zv}(b) shows the $Z-v_{par}$ correlation for events that we mostly ascribe to QP break-up. Indeed four loci are mainly filled: according to the quadrants defined by the dashed lines, we verified that $BF$s with $Z>10$ emitted at $v_{par}>70\,$mm/ns ``Heavy-Fast'') are mainly correlated with lighter $BF$s with $v_{par}<70\,$mm/ns ``Light-Slow''); $BF$s with $Z>10$ emitted at $v_{par}<70\,$mm/ns (``Heavy-Slow'') are correlated with lighter $BF$s at $v_{par}>70\,$mm/ns (``Light-Fast''). Such observation is compatible with the well known QP break-up scenario~\cite{bib:casini93, bib:stefanini95, bib:defilippo12_timescale, bib:jedele17}. We can strenghten this selection by means of the correlation between the relative angle of the two detected fragments $\theta_{rel}$ (in the system center of mass)  and their relative velocity $v_{rel}$. Indeed, in such a correlation QP break-up events settle at low $\theta_{rel}$ and at a $v_{rel}$ compatible with that of a Coulomb-driven split~\cite{bib:piantelli19_isofazia}. On the contrary, coincidence between QP and QT lies at $\theta_{rel}$ values close to 180$^{\circ}$. Results are shown in Fig.~\ref{fig:hsel_zv}(c). Consequently, the QP break-up (\qpb{}) channel events are selected requiring $M_{BF}=2$ and the two $BF$s in the phase-space region within the red contour of Fig.~\ref{fig:hsel_zv}(c). In addition, we require that the total charge of the two $BF$s is within the aforementioned defined QP charge range (\ie{} $12-22$). Events selected as described are the 1.5\% of the total events (corresponding to the 75\% of the $M_{BF}=2$ sample).

\subsection{Evaporative and break-up channel characterization}

Since both selected channels could contain partially detected 
events of higher multiplicity, the study of their gross properties is mandatory in order to validate the selections. For such purpose, we exploited the AMD+GEMINI++ model, which has shown to be able to reproduce the gross properties of heavy-ion collisions in a large range of ions and bombarding energies~\cite{bib:tian17_cc, bib:tian18_cc, bib:piantelli19_fiasco, bib:piantelli19_isofazia}.

Preliminary, the percentages predicted by the simulation for \qpe{} and \qpb{} events are 65\% and 1.5\%, \ie{} in agreement with the values observed in the experimental dataset. Moreover, the amount of \qpb{} events within the \qpe{} selection is below 2\% (due to the limited geometrical acceptance), thus allowing to go further in the event characterization. 

The measured distributions of the \qpr{} charge, parallel velocity in the laboratory frame, and diffusion angle in the system center of mass are reported in Fig.~\ref{fig:hsel_gross}(a, b, c), for the \caa{}+\caa{} reaction, respectively; results for the \ca{}+\ca{} reaction are shown in fig Fig.~\ref{fig:hsel_gross}(d, e, f). 
Both \qpe{} and \qpb{} channels are shown. 
Each distribution is normalized to unity for a better shape comparison with the model prediction; \qpb{} distributions are further scaled by a factor 0.5 for sake of clarity. We underline that in the \qpb{} channel, the QP is reconstructed from the two detected $BF$s.

For the experimental case, we observe that both the parallel velocity ($v_{par}$) and the diffusion angle ($\theta_{cm}$) show typical features of binary dissipative collisions. Indeed, for \qpe{} events extends downwards starting from beam velocity, while the $\theta_{cm}$ is peaked at angles slightly larger than the grazing angle~\cite{bib:piantelli19_fiasco}. 
Similar characteristics are also found in the \qpb{} distributions. However, some differences arise. The larger widths of the three distributions observed for \qpb{} are consistent with the expected broader phase-space region for \qpb{}, and the laboratory velocity tends to be on average smaller than for \qpe{} events. The AMD+GEMINI++ simulation is in global agreement with the observed distributions, as also shown in a recent investigation on Kr+Ca reactions at 35\amev{} with four FAZIA blocks~\cite{bib:piantelli19_isofazia, bib:piantelli20_isofazia2}. We remind that the simulation was subjected to the same constraints as the experimental data. For the \qpe{} channel, the simulation follows the experimental trend, especially in the \ca{}+\ca{} reaction, while some slight discrepancies appear for the \caa{}+\caa{} reactions. 
Such differences could be related to a different dissipation degree between the experimental and the simulated data. Indeed, the model seems to favor more dissipative events, \ie{} lighter \qpr{} (panel a), lower parallel velocity (panel b), and with larger diffusion angle. Similar findings have been found also in the Kr+Ca comparison with the AMD+GEMINI++ predictions~\cite{bib:piantelli19_isofazia, bib:piantelli20_isofazia2}. 

As a final note of this section we observe that the QP distributions for the asymmetric \caa{}+\ca{} system are very similar to those of the symmetric \caa{}+\caa{} case (fig.~\ref{fig:hsel_gross}(a,b,c)). This is reasonable since we are observing very similar \qpr{} and none of the characteristics shown so far take into account the detailed isotopic composition of the ejectiles. 
In conclusion, as also in the recently investigated Kr+Ca reactions with four FAZIA blocks\cite{bib:piantelli19_isofazia, bib:piantelli20_isofazia2}, the AMD+GEMINI++ simulation offers a reasonable description of both the \qpe{} and \qpb{} channels, thus confirming the validity of the adopted selection criteria.

\section{Reaction dissipation and centrality}
\label{sec:dissipation}

\begin{figure*}
\centering 
\includegraphics[width=0.75\textwidth]{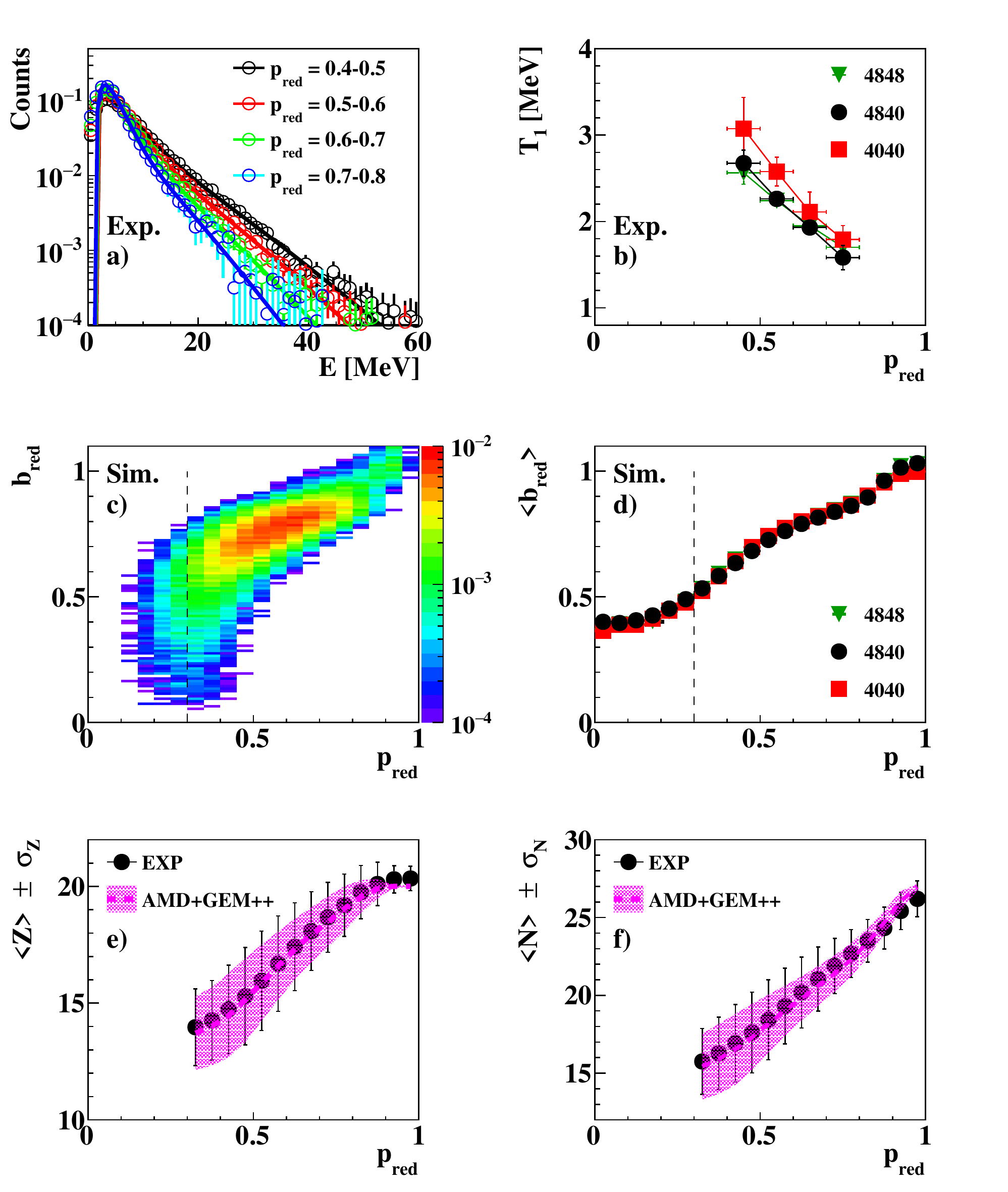} 
\caption{(Color online) Experimental data for the \caa{}+\ca{} system: panel a) proton kinetic energy spectra in the \qpr{} frame for different bins of \pred{}, normalized to unitary area; panel b) average kinetic temperature $T_{1}$ as a function of \pred{} extracted from the Maxwellian fit (shown in panel a)) for the proton kinetic energy spectra; the results for the three systems are presented with symbols according to the legend; only statistical errors of the fit are shown.
Simulated data (filtered AMD+GEMINI++ simulation): panel c) reduced impact parameter $b_{red}$ \vs{} reduced momentum $p_{red}$; panel d) average reduced impact parameter \bredb{} \vs{} \pred{} for each system. Comparison between experimental and simulated data for the \caa{}+\ca{} system: panel e) average \qpr{} charge and sigma of the charge distribution as a function of \pred{}; panel f) same as e) for the neutron number distribution. Symbols according to the legend.} 
\label{fig:h_best} 
\end{figure*} 

In this section, we aim at extracting an experimental 
observable which can be used to order the events as a function 
of the reaction dissipation, to quantify the isospin diffusion from peripheral to more central events. The chosen observable is based on the momentum of the detected (or reconstructed) \qpr{}. We define the reduced momentum (\pred{}), defined as $p_{red} = \left( \frac{p^{QP}_{par}}{p^{beam}} \right)_{cm}$, i.e. the QP remnant (or reconstructed) parallel momentum ($p^{QP}_{par}$) normalized to the beam momentum ($p^{beam}$), both of them in the c.m. frame. 

We first verify, for the experimental data, that the reduced momentum scales as a function of the reaction dissipation. We report the results from the \caa{}+\ca{} reaction as a representative case. We focus on the \qpe{} channels since no significant amount of LCPs are detected in the \qpb{} channel due to the limited angular setup. For such purpose we exploited the LCPs forward emitted with respect to the \qpr{}, that more reliably can be attributed to the QP decay, being less affected by other contributions. However, in this phase-space other contributions could be present, as LCPs associate to pre-equilibrium emissions. One expects that the LCP coming from the statistical decay of the QP present a Maxwellian-like kinetic energy spectra: the apparent temperature increases with the reaction dissipation. Fig.~\ref{fig:h_best}(a) shows the experimental invariant proton kinetic energy spectra, in the \qpr{} frame, for the \caa{}+\ca{} system: each distribution refers to a different bin of \pred{}, according to the legend, and is normalized to unitary area for better shape comparison. We observe that each distribution presents two slopes, 
corresponding to two apparent temperatures $T_1$ and $T_2$, and this deserves some comments. The \qpr{} is the matching source only for protons that contribute to the low energy tail ($T_1$), i.e. the thermal-part of the distributions~\cite{bib:vient18_thermometry}; the high energy tail ($T_2$) could be due to different mechanism, such as pre-equilibrium emission from the neck~\cite{bib:vient18_thermometry} or from the deformed QP~\cite{bib:piantelli02_fiasco, bib:piantelli07_fiasco, bib:rudolf93_emission}, \ie{} due to protons emitted from different sources. For what is relevant to the present discussion, a two-temperature fit can be used in order to disentangle the thermal part from the non-thermal one, thus obtaining a crude indication on the excitation scale of the QP source. 

The results of the fitting procedure using two
Maxwellian contributions are depicted in Fig.~\ref{fig:h_best}(a), superimposed to the experimental spectra. The values of the fitted parameter $T_1$ are shown in Fig.~\ref{fig:h_best}(b) as a function of \pred{} for all the systems. The obtained $T_1$ scaling as a function of the reduced momentum confirms that, on average, we are indeed selecting collisions with increasing dissipation when \pred{} decreases from 1 to 0.3.

Within the AMD+GEMINI++ model, on the other hand, we can
directly verify the relationship between \pred{} and the reduced impact parameter $b_{red}$ ($b/b_{gr}$). Fig.~\ref{fig:h_best}(c) shows the \bred{} \vs{} \pred{} correlation predicted by the AMD+GEMINI++ simulation, filtered with the 
detector response: the correlation is narrow for peripheral 
collisions and tends to broaden for low \bred{}. 
For this reason, we restrict the following analysis only to the upper-right region indicated by the dashed lines in the figure. Here, the correlation is relatively narrow and permits to reliably explore the range $b_{red} \approx 0.5 - 1$. These findings are quite the same for the three studied Ca reactions as evidenced in Fig.~\ref{fig:h_best}(d) by the evolution of the average reduced impact parameter ($\langle b_{red} \rangle$) as a function of \pred{}. 

Finally, the \qpr{} average charge $\langle Z \rangle$ and the rms width $\sigma$ of the charge distribution are reported as a function of \pred{} in Fig.~\ref{fig:h_best}(e). Panel f) is for the average \qpr{} neutron number distribution $\langle N \rangle$. In particular, the experimental data are shown in black, with the bars indicating the $\pm$1 $\sigma$ values. The model results are in magenta and the $\pm$1 $\sigma$ values are drawn as a contour. As \pred{} decreases, $\langle Z \rangle$ and $\langle N \rangle$ decrease starting from values very close to the projectile ones. The average trends as a function of \pred{} are well reproduced by the simulation and, to a lower extent, also the $\sigma$ of both distributions. The global agreement between the experimental results and simulation strengthens the use of \pred{} as an order variable, in order to explore neutron-proton equilibration as a function of the reaction centrality.

\section{Neutron-proton equilibration: evaporative and break-up channels}
\label{sec:eq}

\begin{figure}
\centering 
\includegraphics[width=1\columnwidth]{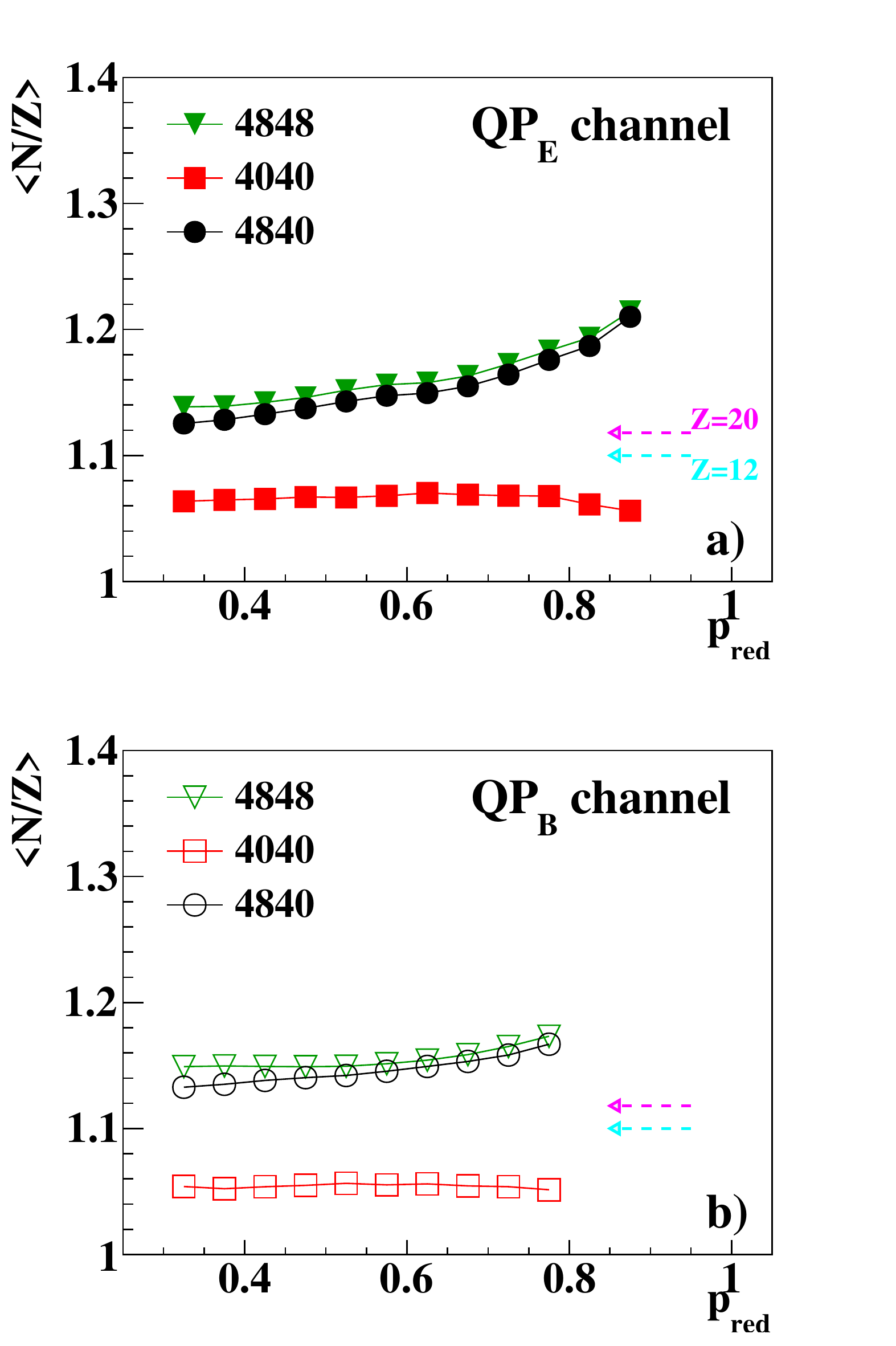} 
\caption{(Color online) Average neutron-proton ratio as a function of \pred{}. Panel a) \qpe{} channel; panel b) \qpb{} channel. Magenta and cyan dashed arrow point out the EAL~\cite{bib:eal} value for Ca and Mg nuclei. Statistical errors are smaller than the marker size. Lines are drawn to guide the eyes.} 
\label{fig:h_nsuz} 
\end{figure} 
\begin{figure}
\centering 
\includegraphics[width=1\columnwidth]{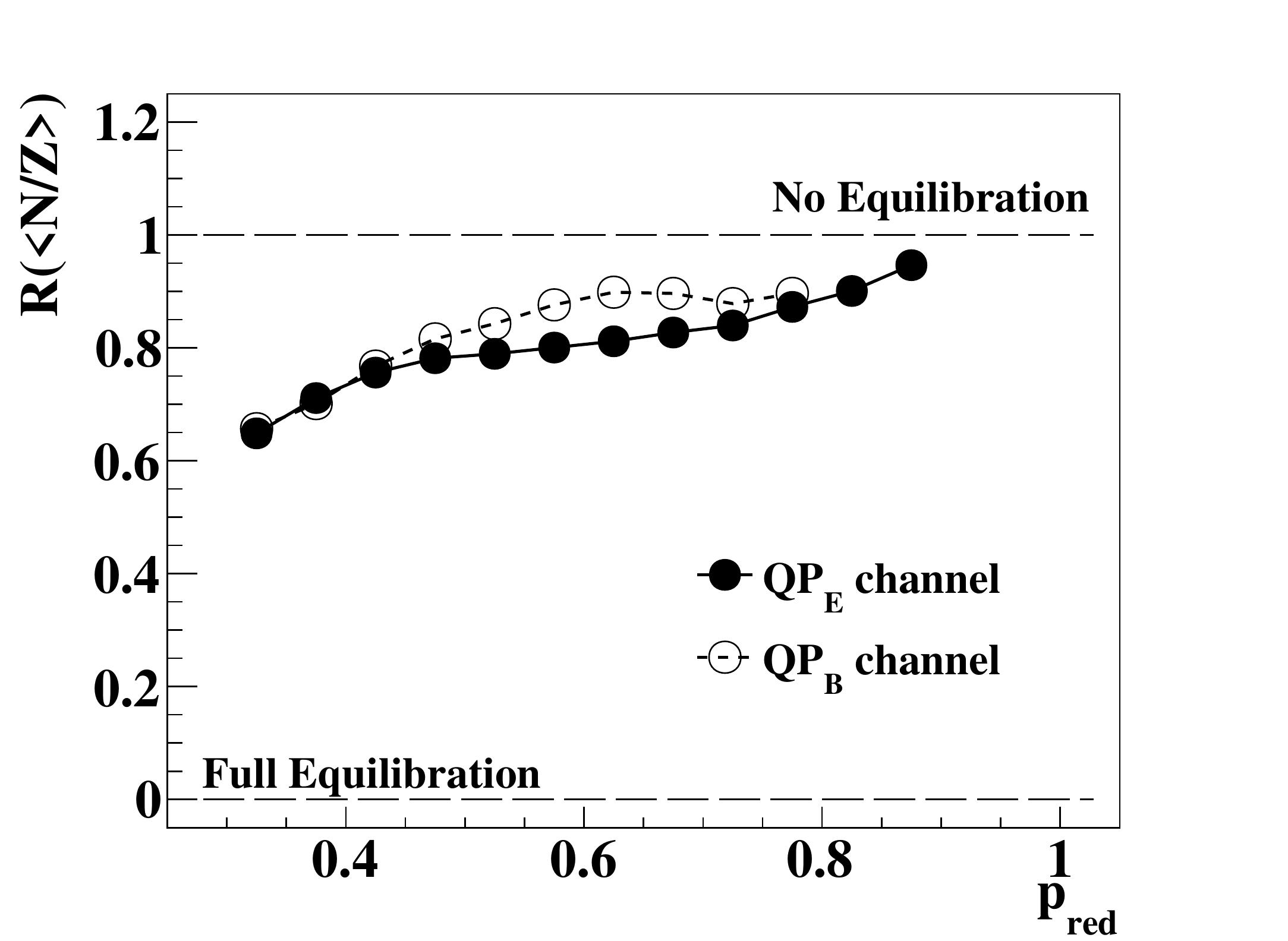} 
\caption{Isospin transport ratio for the \qpe{} and \qpb{} channels as a function of \pred{}. Statistical errors are smaller than the marker size. Symbols according to the legend. Lines are drawn to guide the eyes.} 
\label{fig:h_imba} 
\end{figure} 

The n-p equilibration can be now explored using the average
neutron-proton ratio (\nsuz{}) of the various sources as a function of the reduced momentum. Fig.~\ref{fig:h_nsuz}(a,b)
shows the evolution of \nsuz{} \vs{} \pred{} for the three systems for both the \qpe{} and the \qpb{} channel, respectively. In particular, in the \qpe{} channel, the values refer to the \qpr{}, while in the \qpb{} channel to the reconstructed (from the two $BF$s) QP. For sake of clarity, we remind that the accompanying LCPs and/or IMFs are not taken into account.

As suggested in Sec.~\ref{sec:evselandgross}, we observe that the break-up channel is detectable at lower \pred{}. Apart from this, we observe comparable trends in the two
channels. Namely, the bound neutron abundances of the \caa{} and \ca{} detected (or recontructed) ejectiles are very different as expected, with much larger values for the n-rich case. Such effects are in agreement with studies at lower bombarding energies, mainly dedicated to the investigation of the initial neutron-proton unbalance effects in fusion reactions~\cite{bib:bonnet08_krca, bib:ademar11_krca, bib:pirrone19_krca}. Moreover, the \nsuz{} ratios evolve with dissipation in a different way depending on the initial neutron abundance. 
\caa{} projectiles we observe a sizable decrease of \nsuz{} with centrality, while for the \ca{} case the n-p ratio is essentially constant after a slight increase in peripheral events. 
These different trends can be interpreted in the light of a dominating statistical decay process for n-rich or n-deficient excited nuclei. Indeed, the steep decrease of the average \nsuz{} with respect to the projectile values (1.4 and 1 for the n-rich and n-deficient system, respectively) is mainly due to the statistical decay~\cite{bib:camaiani20_ratio}. As explained in~\cite{bib:eal}, excited nuclei follow an average path in the $N-Z$ plane during the decay and, with increasing initial excitation, tend to approach a specific region of that plane, called  Evaporation Actractor Line (EAL)~\cite{bib:eal}, 
described by a $N/Z$ ratio, depending on the nuclear size.
In Fig.~\ref{fig:h_nsuz}, just for reference, the EAL $N/Z$ ratios indicated with dashed arrows for ion charges $Z=12,20$ representing relevant values for our QP remnant selection. We see that, with increasing dissipation, \qpr{} from \caa{} and from \ca{} have \nsuz{} values that move towards the EAL predictions, although coming from
different sides.

The comparison between the \nsuz{} of \qpr{} from \caa{} of the symmetric and asymmetric reactions reveals the trend to isospin equilibration. Focusing on the 
\qpe{} case (Fig.~\ref{fig:h_nsuz}(a)), a clear hierarchy is 
observed: a reduced neutron content is detected for the asymmetric
case (black solid circles in fig.~\ref{fig:h_nsuz}(a)) with a gap with respect to the symmetric reference (green solid triangles in fig.~\ref{fig:h_nsuz}(a)) increasing towards central collisions, as the result of the interaction with a n-deficient partner so that the two colliding nuclei tend to equilibrate their $N/Z$ ratios
~\cite{bib:defilippo12_timescale, bib:barlini13_tele, bib:csym16}.
Remarkably, very similar observations can be repeated for 
the \qpb{} channel, where the same hierarchy and evolution are evident.

In order to more quantitatively establish the isospin equilibration process we show in Fig.~\ref{fig:h_imba} the isospin transport ratio $R(X)$ built with $X=\langle N/Z \rangle$ (Eq.\ref{eq:ratio}) 
as a function of the reaction dissipation represented by \pred{}.
Concerning the evaporative channel, we observe the expected trend. 
The equilibration degree smoothly and monotonically evolves from $R\approx1$ for $p_{red} \approx 1$ to $R \approx 0.6$ for $p_{red}\approx 0.3$, 
which, according to the AMD average prediction (Fig.~\ref{fig:h_best}(d)), corresponds to a range of centrality $\langle b_{red} \rangle \in [1,0.5]$.
Also the experimental result
for the same Ca+Ca collision~\cite{bib:phd_quentin, bib:boisjoli12_caca,
bib:wigg12_caca} obtained with the INDRA+VAMOS
experimental apparatus points out in this direction~: the n-p equilibration for such experiment is
compatible, as discussed in~\cite{bib:phd_camaiani}, with that here reported. 
The isospin diffusion sets in for the asymmetric reactions and makes the QP and QT to approach a common $N/Z$ values.
Since the QP size selection is somewhat arbitrary (Sec.~\ref{fig:hsel_gross}), we tested the
result by changing the adopted QP charge range. In particular, we
increased and decreased the lower limit of two units with respect to our previous ``standard'' ($Z=12$) value (as done in~\cite{bib:galichet09_qprebuild,
bib:galanopoulos10_fission}), taking into account other reasonable choices reported in the literature. For instance Ref.~\cite{bib:theriault05}
fixes as a lower limit of the QP charge the 36\% of the projectile charge.
By using the ranges $Z^{QP}\in[14, 22]$ or $Z^{QP}\in[10, 22]$, we
found that the trend of $R$ is negligibly affected in the studied range of \pred{}~\cite{bib:phd_camaiani}. 

An important point of this work, as anticipated, is the access to the
isospin diffusion process looking at the \qpb{} channel, in a manner that - to our knowledge - has not been yet attempted before. In Fig.~\ref{fig:h_imba} the open dots show the \nsuz{} for QP reconstructed from the break-up fragments. 
As a first comment we can say that the general trend is the same, with slight differences: for the \qpb{} we find a weak process at least for the less dissipative accessible bins. 
It is very difficult to judge and conclude about these small
differences which, in any case, are out of statistical errors. 
Such an observation suggests a heavier primary source in the \qpb{} channel, which can lead the system to a lower n-p equilibration for the most explored peripheral events. For instance, the average charge and neutron number of the reconstructed QP in the \qpb{} channel are on average 2 units larger than the values of the \qpr{} in the evaporative channel. 
On the other hand, the differences can be also related to subtle effects associated to the different evaporation paths followed by the excited break-up fragments (before and after the split) with respect to the case without break-up. 

Such a topic will be further investigate in the INDRA+FAZIA experimental campaign at GANIL, thus combining with the isotopic capabilities of the FAZIA multi-telescope array the large angular coverage of the INDRA detector, in order to more precisely select the reaction centrality. Here, we can only conclude that this roughly common trend of the two geometrical loci in fig.\ref{fig:h_imba} suggests that, irrespective of the final state channel, the isospin
diffusion acts in a similar way. In other words, it appears that the isospin equilibration process acts before any de-excitation process.
This observation is rather in line with some old results~\cite{bib:planeta}
for lower energy collisions. There, a general conclusion was
suggested that the n-p degree of freedom tends to relax rather quickly during the interaction. 
The complete equilibrium could be reached only for
rather central impacts, not accessible here according to the AMD centrality estimation of Fig.~\ref{fig:h_best}(c,d), associated with relatively long interaction times.

\section{Neutron-proton equilibration: comparison with the simulation}
\label{sec:amd}

In this section, we aim at comparing the isospin evolution  extracted
from experimental data with that predicted by the transport model
AMD, coupled with GEMINI++ as an afterburner. We will focus on the evaporative channel, as it corresponds to 65\% of the collected
data. The break-up channel is experimentally around 35 times less abundant and since also the model predicts a similar event partition, the simulation statistics results to be to low for a reliable comparison. For sake of clarity, we remind that the simulated data have been treated as the experimental ones.

\begin{figure}
\centering 
\includegraphics[width=1\columnwidth]{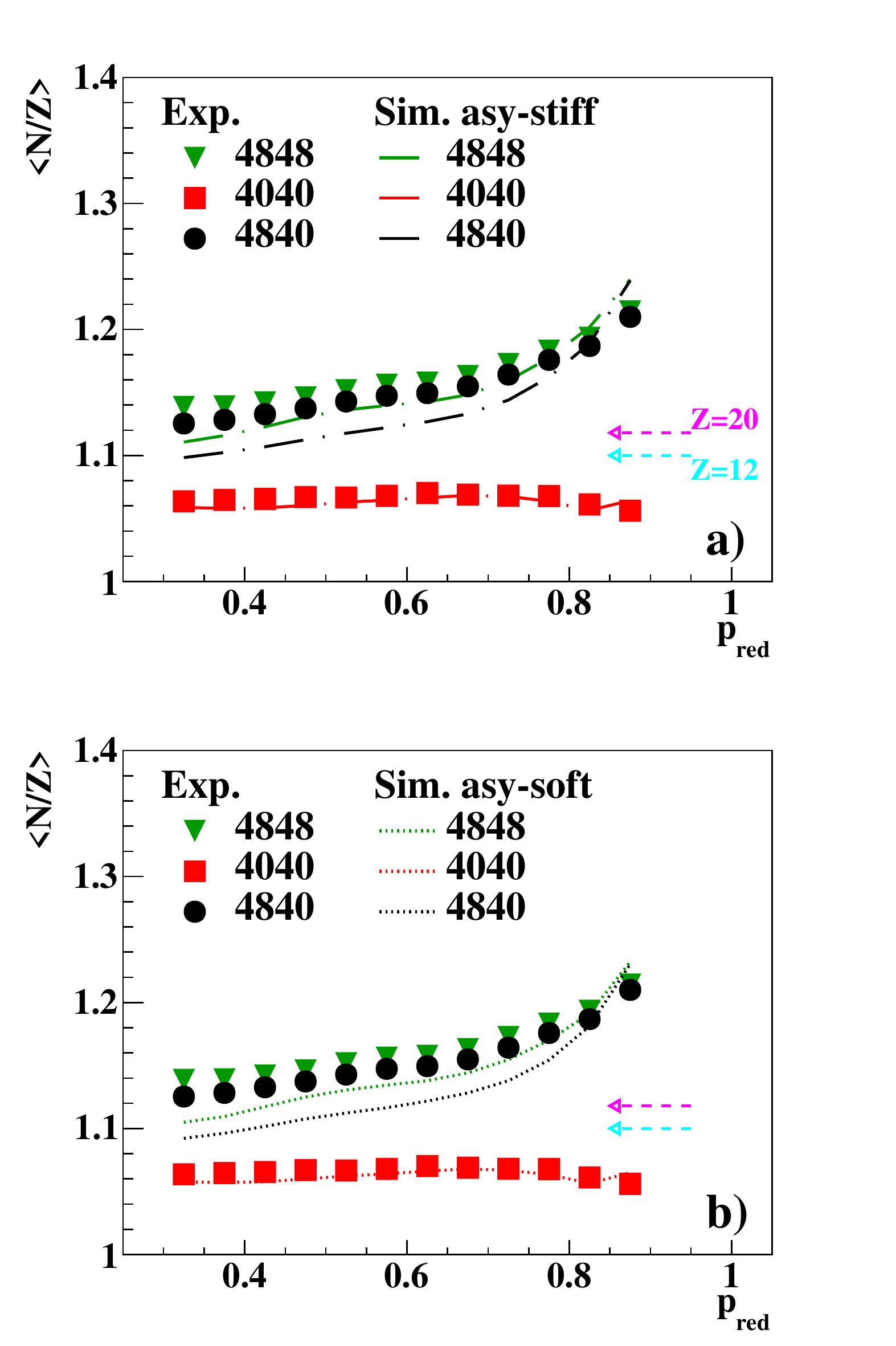} 
\caption{Comparison of the measured average neutron-proton
ratio for the \qpe{} channel as a function of \pred{} with the
AMD+GEMINI++ simulation. Panel a) AMD asym-stiff parametrization;
panel b) AMD asym-soft parametrization. Magenta and cyan dashed
arrow point out the EAL~\cite{bib:eal} for relevant nuclei. Symbols according to the legend.  
Statistical errors are smaller than the marker size (line width).} 
\label{fig:h_nsuz_amd} 
\end{figure}

Fig.\ref{fig:h_nsuz_amd}(a,b) shows the simulated \nsuz{} \vs{} \pred{} trend (lines), compared with that obtained experimentally (same points of Fig.~\ref{fig:h_nsuz}(a)) for the asym-stiff and asym-soft parametrization of
the symmetry energy, respectively. As for the experimental data, we
observe the clear hierarchy among the three systems, and the tendency to approach \nsuz{}
values around the EAL loci (magenta and cyan arrows for $Z=20$ and
$Z=12$, respectively) with increasing dissipation. The
agreement with the \nsuz{} of the \ca{} data is excellent while, as
noticed for the gross properties of the \qpr{}
(see. Sec.~\ref{fig:hsel_gross}), there are some differences for the
\caa{} case. Weak differences between the two calculations can be seen, in particular, the asym-stiff
choice predicts a more neutron-rich \qpr{} with respect to the
asym-soft one, as expected~\cite{bib:baran05_eos, bib:baran05_transport}.

\begin{figure}
\centering 
\includegraphics[width=1\columnwidth]{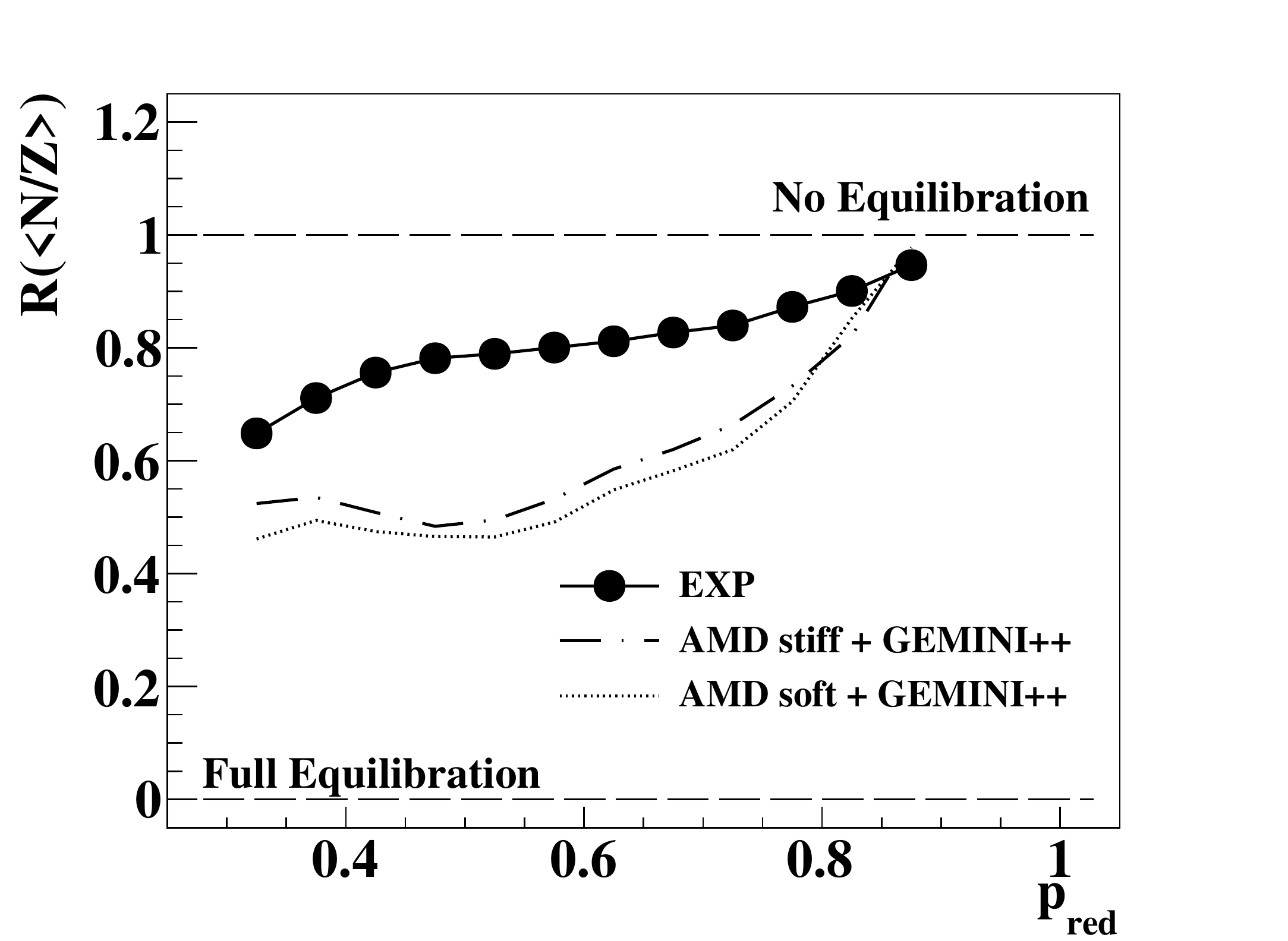} 
\caption{Comparison of the isospin transport ratio for the \qpe{}
channel as a function of \pred{} between the experimental results and the AMD+GEMINI++ simulation, using asym-stiff and asym-soft parametrizations. Statistical errors are smaller than the marker size (line width).} 
\label{fig:h_imba_amd} 
\end{figure}

The corresponding isospin transport ratio are shown in Fig.~\ref{fig:h_imba_amd} as a function of \pred{}, with dot-dashed and dotted line for the asym-stiff and asym-soft parametrizations, respectively. We first underline that the $R$ variable depends on the gap between the asymmetric and the symmetric references. 
The way how the gap evolves \textit{vs.} \pred{} dictates the shape of the R as a function of the dissipation, thus a precise reproduction of the \nsuz{} values is not mandatory. 
However, Fig~\ref{fig:h_imba_amd} shows a sizable disagreement between experiment and model predictions concerning the isospin diffusion process. In particular, the model predicts an initial fast relaxation followed by a slower trend, whereas the experiment suggests a smoother evolution. As for the asym-stiffness, we can see that the very small differences in the
two model results for \nsuz{} give a quite small gap in the
equilibration degree; however, as expected, the asym-soft assumption
slightly favors the isospin relaxation. 


Some comments and arguments on the possible origin of the observed
disagreement are in order.
A first comment deals with the role of the emissions from the primary
QP, i.e. the fragment emerging just at the end of the interaction
which we would like to access in order to measure the isospin
diffusion. Indeed, any particle or fragment emission before the
detection perturbs the final isotopic distribution.
One can thus wonder if the found disagreement is related to a
partially wrong description of the dynamics (reaction times and/or
nuclear potential terms ruling the isospin transfer) or to a somehow
wrong evaporation scheme. In this respect, we must stress
that isospin transport ratio has been introduced~\cite{bib:rami00_imbalance, bib:betty04_isoscaling}  just to 
bypass any perturbation which introduces a linear transformation of
the isospin variable in use (Eq.~\ref{eq:ratio}). Such behavior has
been recently investigated in a specific
work~\cite{bib:camaiani20_ratio}, in a full model framework, for the systems here discussed. In this paper one 
demonstrates, by means of the AMD simulation coupled with statistical models, that the
charge equilibration process measured via isospin transport ratio is
indeed affected
by perturbations introduced by the dynamical and statistical emissions
from the fragments after their separation. In
particular, the statistical emission (described by \gem{} code) tends to introduce non-linear
spurious distortions at low excitation energies (where structure effects are well known to affect the particle emission~\cite{bib:camaiani18, bib:more19, bib:bruno19}), i.e. for large impact parameters, while the distortion becomes smoother
and linear with increasing excitation. 
Instead, at least for the considered systems, the contribution of emissions occurring
during the interaction phases and predicted by the AMD model, increases with centrality but remains relatively scarce and negligibly affects the $R$ variable.
As a consequence, we checked that despite
some residual distortions related to emissions, the variable $R$ is robust and keeps memory of the primary isospin history; this suggests that the observed discrepancy between measured and predicted $R$ can be safely ascribed to the dynamical modelization.

By analyzing the evolving output of the model, we can access to the end of the projectile-target interaction phase (labeled as $t_{DIC}$), by means of the procedure described in refs.~\cite{bib:camaiani20_ratio, bib:piantelli19_isofazia}. 
In order to pin down the mechanism responsible for the observe discrepancies with experiment, we applied some special conditions on the analyzed events, as follows.
\begin{figure}
\centering 
\includegraphics[width=1\columnwidth]{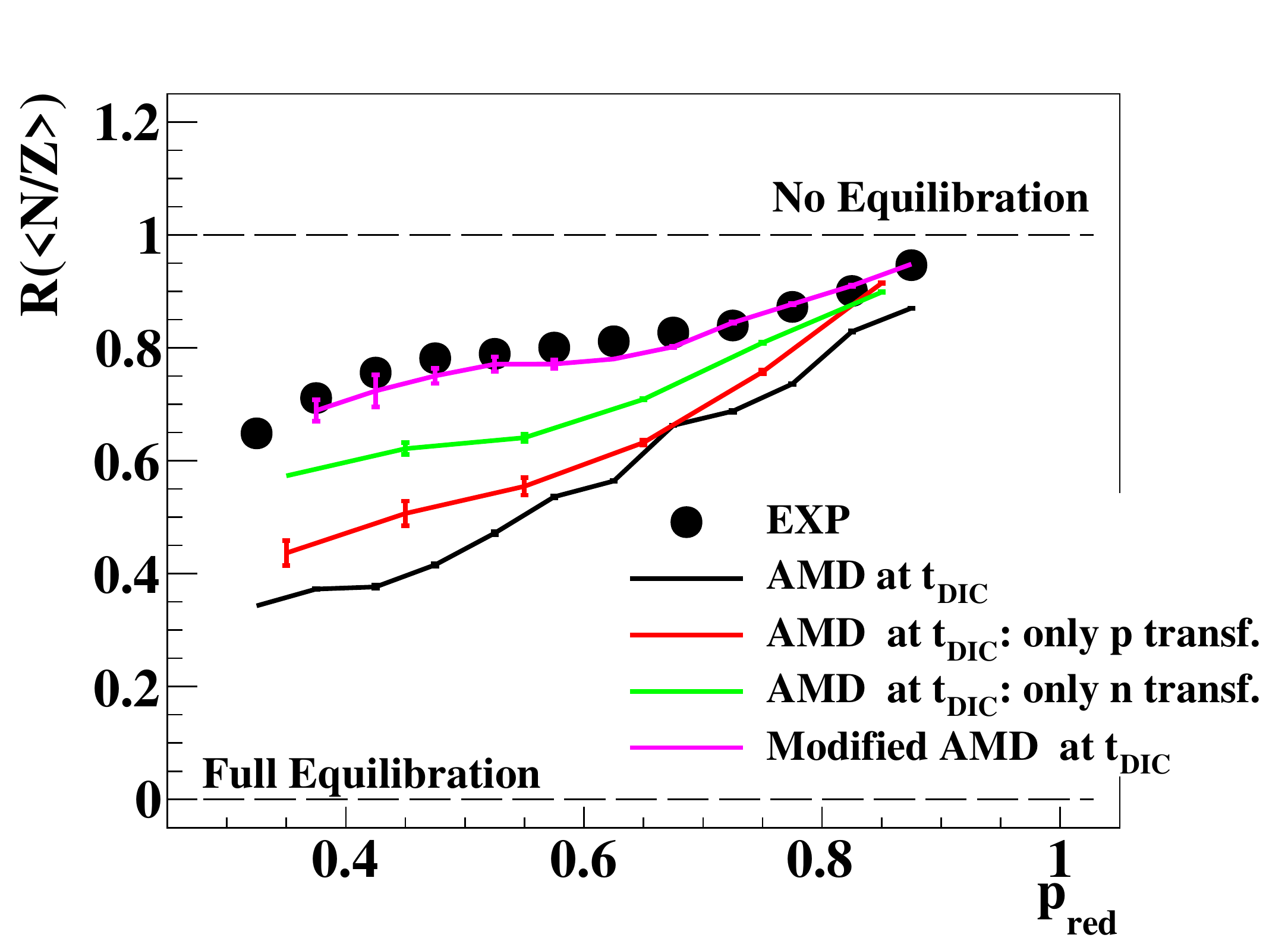} 
\caption{(Color online) Comparison of the experimental isospin transport ratio with the equilibration obtained at the projectile-target separation time ($t_{DIC}$), for the asym-stiff parametrization; the isospin transport ratios only due to a net charge (red line) and neutron (green line) number change are shown. The equilibration obtained after a re-scaling of the proton and neutron transfer probabilities is shown with magenta line. See text for details. Errors are statistical.}
\label{fig:h_tdic} 
\end{figure} 
The n-p equilibration obtained at $t_{DIC}$, for the asym-stiff simulation, is shown in Fig.~\ref{fig:h_tdic} as a black line; for sake of comparison also the experimental trend of Fig.~\ref{fig:h_imba_amd} is here reported. For each system (\ie{} the asymmetric and the symmetric references), we start allowing only the net neutron transfers (green line): this corresponds to retain only the reaction channels where the QP emerges as a Ca isotope. Vice versa, we allow only the net proton exchanges (red line), \ie{} events where the QP retains the neutron number of the projectile. As expected,  limiting the n-p exchange produces a lower equilibration. More interesting, we observe that the equilibration obtained via only charge change lies close to the total one, pointing out to an important role of the p transfers in the isospin equilibration mechanism. This can be quantitatively understood keeping into account that, in order to restore the $N/Z$ unbalance, a p transfer is more effective than a n transfer, since the former counts as 1/20 whereas the latter as 1/28. 

Starting from the indication that the nucleon transfer in AMD may be too frequent, we now aim at quantifying the degree of the overestimation of the transfer probability.  We introduce a multiplying factor ($f$), depending on the net number of transferred neutrons and protons, $\Delta_n$ and $\Delta_p$ respectively. Assuming that nucleon transfers in the same event are  independent of each other, we modelled a parametrization as:
$f=\alpha^{|\Delta_n|}\beta^{|\Delta_p|}$, where $\alpha$ and $\beta$ are parameters to suppress (or enhance) the net transfer probability of single neutrons and single protons, respectively. 
The probability of the non-transfer channel (at $t_{dic}$) is adjusted for the total probability conservation.
For each system, we then proceed to classify the various channels as function on the net p/n changes at $t_{DIC}$: we modify these initial populations via a change of the $(\alpha,\beta)$ pair and thus
obtain different average isospin values. The isospin transport ratio is then computed via eq.(\ref{eq:ratio}), adopting the $\langle N/Z \rangle (\alpha,\beta)$ as $X$ variable ($R_{AMD}(\alpha,\beta)$). 
The parameters $\alpha$ and $\beta$ have been selected by means of fit procedure on the experimental data $R_{exp}$.
Specifically, we looked for the minimum of a $\mathbf{M}^{2}$ variable defines as follows:
\begin{equation}
\mathbf{M}^{2} = \sum_{i=0}^{N} \frac{\left[R_{exp}^{i}-R_{AMD}^{i}(\alpha,\beta)\right]^2}{\sigma_{exp}^{2}(i)+\sigma_{AMD}^2(i)},
\end{equation}
where $R_{exp}^i$ and $R_{AMD}^i$ are the values of the experimental and simulated $R$ at the $i$th point along the \pred{} axis; $\sigma_{exp}^2(i)$   and $\sigma_{AMD}^2(i)$ the statistical errors of each point. 
The fitted values of the parameters are: 
$\alpha = 0.60 \pm 0.05$, $\beta = 0.3 \pm 0.1$. The equilibration degree obtained for such values is shown in Fig.~\ref{fig:h_tdic} with magenta line (Modified AMD), which follows the experimental trend proving the satisfactory quality of the fit. This show that the nucleon transfer is overestimated in AMD by about a factor of two. Moreover, it is likely that proton transfer is more overestimated than neutron transfer. 

In conclusion, this first attempt to compare the n-p equilibration measured via the isospin transport ratio built from the $\langle N/Z \rangle$ of the \qpr{} has shown a faster equilibration of the model prediction with respect to that observed in the experimental sample. Such discrepancy can be recovered acting on the transfer probability, reducing it approximately of a factor two. It is not easy to identify a reason behind this problem, as many factors could contribute to it, e.g. the nucleon-nucleon cross section or the nucleon effective masses or their interplay. For instance, a simple variation of the screening paramenter $y$ of the nucleon-nucleon cross section from $y=0.42$ up to the free nucleon-nucleon cross section did not produce significant variations of the isospin transport ratio. Such topics will be investigated in future works.

\section{Summary and Conclusion}
\label{sec:concl}

In this paper, we have presented the experimental results of an
experiment dedicated to the investigation of the n-p equilibration in
\caa{}+\ca{} semi-peripheral reactions at 35\amev{}, performed with
four blocks of the \faz{} multi-telescope array at the INFN-LNS. For
the first time, thanks to the \faz{} identification performances
coupled to its good granularity, we could study the isospin relaxation for the two main QP decay channels, the evaporative and the break-up
one.

The equilibration trend has been investigated by means of the isospin
transport ratio, which which improves the sensitivity to the effect sought after and normalizes the mixed system evolution with the
limiting values of the symmetric reactions \caa{}+\caa{} and \ca{}+\ca{}, investigated under the same
experimental conditions. Despite the relatively small coverage
of the setup ($2-8^{\circ}$ in the laboratory frame), the main
achievements have been proved not to be strongly affected by the
apparatus response: indeed we focus on the QP phase-space for which we
have reasonable acceptance. We have introduced a reaction dissipation estimator ($p_{red}$), which has been linked with 
the reaction centrality by means of the model.

The results reported in this paper are the following.
As expected, the relaxation of the isospin degree of freedom has been
observed in the \caa{}+\ca{}, via the use of the isospin transport
ratio of the average neutron-proton ratio ($\langle N/Z \rangle$)
of QP remnants.

The comparative analysis of the QP evaporative and break-up channels
has shown the typical signature of the isospin diffusion: as the reaction centrality
increases, the system evolves to restore the charge equilibrium. 
The similarity of the behavior for the two channels suggests
a comparable dynamical evolution before the decay, whatever it is.
Specifically, this is consistent with an isospin exchange mechanism
that acts on a similar timescale (that of the interaction phase) shorter than the evaporation cascade or the QP split phase.

Concerning the comparison with the AMD model coupled with the GEMINI++ statistical code, we observed that the model globally reproduces the main features of the QP in both the evaporative and break-up
channels; the agreement is better for the QP evaporation channel than
for the break-up one, where the model produces lighter and
slower fragments than the measured ones. Also, the agreement is quite good
for the \ca{} system while for the \caa{} reactions it less nicely
reproduces the QP data. The detailed isospin distributions of the final (post-evaporative) fragments are, again, less well reproduced for the n-rich systems; for the \ca{} reaction the comparison is excellent.

The main difference between measured and model data is observed in the evolution towards the charge equilibration for the evaporative exit channel. The model predicts a faster  relaxation of the initial neutron-proton unbalance with respect to the experiment. This discrepancy seems to be associated with an overestimated probability of nucleon transfers, mainly and more specifically for the protons: in particular a reduction of about a factor two accounts for the experimental path. However, a deeper investigation on this point is in program. In this respect we plan to extend the analysis of this paper to the data obtained by the first recent INDRA-FAZIA experiment on Ni+Ni reactions at comparable energies. Here, we have the almost full isotopic identification of QP ejectiles coupled with a much larger acceptance, allowing to adopt and cross-check several variables, several variables, to extend the analysis to the full panel of exit channels, and to more precisely select the reaction centrality.

\begin{acknowledgements}
This work required the use of a lot of computation
time for the production of the simulated data. We would like to thank the GARR Consortium for the kind use of the cloud computing infrastructure on the platform cloud.garr.it. We would like to thank also the INFN-CNAF for the use of its cloud computing infrastructure. A. Ono was supported by JSPS KAK-ENHI Grant No. JP17K05432. This work was also supported by the National Research Foundation of Korea (NRF) (Grant No. 2018R1A5A1025563).
\end{acknowledgements}

\bibliography{./biblio.bib} 

\end{document}